\documentclass[aps,prd,groupedaddress,12pt,tightenlines,nofootinbib]{revtex4}
\usepackage{amsmath,amssymb,amsthm}

\newtheorem{Theorem}{Theorem}

\newtheorem{Lemma}{Lemma}

\def\refa#1{(\ref{#1})}

\def\ra{\rightarrow}

\def\C{\mathcal{C}}
\def\N{\mathbb{N}}
\def\R{\mathbb{R}}
\def\RTX{\R_+^{1+3}}
\def\Linf{{L^\infty}}
\def\Linfx#1{{L^\infty_{#1}}}
\def\Linftx#1{{L^\infty_{1,#1}}}
\def\Linftau#1#2{{L^\infty_{#1,#2}}}
\def\LinfX{{L^\infty(\R^3)}}
\def\LinfTX{{L^\infty(\RTX)}}

\def\<{\langle}
\def\>{\rangle}
\def\norm#1{\<#1\>}
\def\nnorm#1{(1+|#1|)} 
\def\w#1{\widetilde{#1}}

\def\d{\partial}

\def\eps{\varepsilon}
\def\la{\lambda}

\begin{document}

\title{Linear and nonlinear tails I: \\general results and perturbation theory}

\author{Nikodem Szpak}
\affiliation{Max-Planck-Institut f{\"u}r
Gravitationsphysik, Albert-Einstein-Institut, Golm, Germany}
\date{\today}

\begin{abstract}
For nonlinear wave equations with a potential term we prove pointwise space-time decay estimates and develop a perturbation theory for small initial data. We show that the perturbation series has a positive convergence radius by a method which reduces the wave equation to an algebraic one.
We demonstrate that already first and second perturbation orders, satisfying linear equations, can provide precise information about the decay of the full solution to the nonlinear wave equation.
In a forthcoming publication (part II) we address the issue of optimal decay estimates and precise asymptotics under spherical symmetry where the perturbation equations can be solved almost exactly. 
\end{abstract}


\maketitle

\section{Introduction}

It is a well known fact that the presence of a long-range potential term (power-law decay at spatial infinity) in the wave equation violates the Huygens principle and gives rise to a late-time tail in the solution with a power-law decay (both powers are related) \cite{Strauss-T, NS-WaveDecay}. It is not so well known that nonlinear terms like $u^p$ cause the same effect. 
We study equations where both these effects are present and give pointwise decay estimates on the solutions. Further, we develop perturbation theory for these equations and by its means argue that presented estimates give optimal decay rates at late times. A rigorous proof of this fact will appear in a following publication \cite{NS-PB_Tails} (part II).

We consider linear and nonlinear wave equations with a potential term of the general form
\begin{equation} \label{wave-eq}
   \Box u + V u = F(u)
\end{equation}
in 3 spatial dimensions, i.e. $u:(t,x)\in \R_+\times \R^3\equiv\RTX\ra\R$, and solve the initial value problem with
\begin{equation} \label{init-data}
  u(0,x)=f(x),\qquad \d_t u(0,x)=g(x).
\end{equation}
First, we construct an iteration scheme and show its convergence in a weighted space-time $\Linf$-norm what reproduces the decay estimate from \cite{Strauss-T, NS-WaveDecay}
\begin{equation*}
  |u(t,x)| \leq \frac{C}{(1+t+|x|)(1+|t-|x||)^{q-1}} \qquad \forall (t,x)\in\RTX
\end{equation*}
with $q:=\min(m-1,k,p-1)$ provided the potential $V$ and the initial data $f,g$ satisfy pointwise bounds
\begin{equation*} 
  |V(x)| \leq \frac{V_0}{\nnorm{x}^k},\quad k>2
\end{equation*}
\begin{equation*} 
  |f(x)| \leq \frac{f_0}{\nnorm{x}^{m-1}}, \quad 
  |\nabla f(x)| \leq \frac{f_1}{\nnorm{x}^m}, \quad 
  |g(x)| \leq \frac{g_0}{\nnorm{x}^m},\quad m>3.
\end{equation*}
with small $V_0, f_0, f_1, f_2$ and the nonlinearity is analytic and satisfies for $p>1+\sqrt{2}$
\begin{equation*}
  |F(u)|\leq F_1 |u|^p,\quad |F(u)-F(v)|\leq F_2 |u-v| \max(|u|,|v|)^{p-1}\quad \text{for } |u|, |v|<1.
\end{equation*}

Next, we construct a perturbation series representing the solution $u$ and prove its convergence (with finite convergence radius) in the same weighted space-time $\Linf$-norm. It implies  pointwise convergence in $\RTX$ what allows us to control the decay at every perturbation order and obtain estimate on the remainder of the perturbation series for any order. Finally, if we can show that at some perturbation level our decay estimate is optimal, i.e. we know the true asymptotics for late times (what is not very difficult because the perturbation equations are linear) then we immediately know the asymptotics of $u$. It is the same as that of the given perturbation order because all higher terms in the perturbation series, summed up, are too small to be able to modify the asymptotics. The issue of optimal decay estimates and precise asymptotics compared with numerical results will be addressed in a forthcoming publication \cite{NS-PB_Tails} which will be focused on spherical symmetry where the perturbation equations can be solved almost exactly.

The proof of convergence of the perturbation series is essential for justifying the perturbation scheme as a rigorous approximation and being able to provide exact decay rates. We show it by relating the (inverted) wave equation
\begin{equation*}
  u = \Box^{-1} F(u) - \Box^{-1}(Vu) + \eps I(f,g)
\end{equation*}
(where $\eps I(f,g)$ stands for initial data contribution to the solution of the free wave equation $\Box u=0$) to an algebraic equation of a similar form
\begin{equation*}
  W = C \w{F}(W) + \delta W + \eps D,
\end{equation*}
($\w{F}$ is obtained from $F$ by transformation of its Taylor series), which arises from comparison of the perturbation schemes for both problems. We make an interesting observation that the nonlinear wave equation has a solution $u(\eps)$ analytic in $\eps$, and hence representable by a convergent series in $\eps$, if the same holds for the solution $W(\eps)$ of the corresponding algebraic equation. The latter, however, is always true when $F(u)$ is analytic at $u=0$ what we assume.

Regarding regularity, we can go a safe way and consider only the classical solutions, i.e. assume $(f,g)\in\C^3(\R^3)\times\C^2(\R^3)$, $V\in\C^2(\R^3)$ and $F\in\C^2(\R)$ and obtain $u\in\C^2(\RTX)$. However, all results remain true also for weak solutions where $(f,g)\in\C^1(\R^3)\times\C^0(\R^3)$, $V\in\C^0(\R^3)$ and $F\in\C^0(\R)$ and we have $u\in\C^0(\RTX)$, because the lemmas \ref{Lem:init-data}-\ref{Lem:decay}, which constitute the main ``engine'' of all estimates, preserve the continuity (see \cite{NS-WaveDecay} for a detailed discussion of the weak solutions).

This paper is organized as follows. It has three main sections addressing the linear wave equation with potential, nonlinear wave equation without and with potential, respectively. The idea is to develop tools for the simplest, linear problem and then to generalize them to the nonlinear situation.
Every section has subsections presenting an iterative and a perturbative approach to the construction of solutions and a discussion of the optimal decay rates. Appendix collects some lemmas used in the proofs, cited from other works.

\subsection*{Notation}


With the symbol $\<x\>:=1+|x|$ we define spatial and space-time weighted-$L^\infty$ norms
\begin{equation*}
  \| f \|_\Linfx{m} := \| \norm{x}^m f(x) \|_\LinfX
\end{equation*}
\begin{equation*}
  \| u \|_\Linftau{q}{p} := \| \norm{t+|x|}^q \norm{t-|x|}^{p-q}  u(t,x)\|_\LinfTX
\end{equation*}
of which we will most frequently use 
\begin{equation*}
  \| u \|_\Linftau{1}{p} := \| \norm{t+|x|} \norm{t-|x|}^{p-1}  u(t,x)\|_\LinfTX.
\end{equation*}
Its finiteness guarantees the decay of $u$ like $1/t$ on the lightcone $t\sim |x|$ and like $1/t^p$ for fixed $x$ as well as $1/|x|^p$ for fixed $t$. 
Note that functions with compact support in $\R^3$ belong to all spaces $\Linfx{m}$ with any $m>0$.

We introduce the following notation for solutions of the wave equations. Let $I_V$ be a linear map from the space of initial data to the space of solutions of the wave equation \refa{wave-eq}-\refa{init-data} with $F(u)=0$, so that $u=I_V(f,g)$. For wave equations with a source term and null initial data
\begin{equation*} 
  \Box u + Vu= F,\qquad u(0,x)=0,\qquad  \d_t u(0,x)=0,
\end{equation*}
let's denote the solutions by $u=L_V(F)$, where $L_V$ is a linear map from the space of source functions to the space of solutions to the above problem. Note that, due to linearity, the solution $u$ of a wave equation with source $F$ and non-vanishing initial data $f,g$ is a sum of these two contributions
\begin{equation*}
  u=L_V(F)+I_V(f,g).
\end{equation*}
Observe that if we put the potential term on the r.h.s. we obtain
\begin{equation*}
  \Box u = -V u + F
\end{equation*}
which, treated as a wave equation without potential (on the l.h.s.), is formally solved by
\begin{equation*}
  u=-L_0(Vu)+L_0(F)+I_0(f,g).
\end{equation*}
Here the solution $u$ appears on both sides what seems to make the formula useless, but it will allow us to formulate various iteration schemes, e.g.
\begin{equation*}
  u_{n+1}=-L_0(Vu_n)+L_0(F(u_n))+I_0(f,g)
\end{equation*}
for which we will prove convergence in suitable $\Linftx{q}$ norms.

Finally, we define constants which arise from estimates proved in \cite{NS-WaveDecay}, improved in \cite{NS-DecayLemma}
\begin{equation*}
  C_m:= \max \left(\frac{9}{2(m-2)}, 5\right),
\end{equation*}
\begin{equation*}
  C_{p,q}:=2+\frac{8}{p-1}+\frac{2}{q-1}.
\end{equation*}
The latter will be referred to as a bound on the allowed strength $V_0$ of the potential. Our purpose is to emphasize that this bound, although not optimal, is finite and not arbitrarily small what is crucial when a potential with a given value $V_0$ is studied (like e.g. in the Regge-Wheeler equation describing waves on Schwarzschild geometry).

\section{Linear case with potential}
First, we consider a linear wave equation
\begin{equation} \label{V:wave-eq}
   \Box u + \la V(x) u = 0
\end{equation}
where $\la>0$ is a small parameter, bounded by some finite constant $C_V>0$ (which will be defined later). We first show that a standard iteration scheme converges for all $\la<C_V$ to a solution in $\Linftx{p}$, i.e. there exists a constant $C$ such that
\begin{equation*}
  |u(t,x)| \leq \frac{C}{\norm{t+|x|}\norm{t-|x|}^{p-1}} \qquad \forall (t,x)\in\R_+\times\R^3
\end{equation*}
with some $p>2$ provided the potential $V$ and the initial data $f,\nabla f,g$ are (at least) continuous and satisfy pointwise bounds
\begin{equation} \label{V-bound}
  |V(x)| \leq \frac{1}{\norm{x}^k},\quad k>2
\end{equation}
and
\begin{equation} \label{fg-bound}
  |f(x)| \leq \frac{f_0}{\norm{x}^{m-1}}, \qquad 
  |\nabla f(x)| \leq \frac{f_1}{\norm{x}^m}, \qquad 
  |g(x)| \leq \frac{g_0}{\norm{x}^m},\qquad m>3.
\end{equation}
Then, we show that a perturbation scheme based on expansion in powers of $\la$ is, due to linearity, equivalent to the iteration scheme and the perturbation series has convergence radius $C_V$.

As next, we show that the lowest order $u_0$ has, in general, a different decay estimate than all higher orders, starting from $u_1$. Finally, we prove that either $u_0$ or $u_1$ gives precise information about the decay rate of the full solution $u$.

\subsection{Iteration}
We define an iteration by
\begin{equation*}
   u_{-1} := 0\\
\end{equation*}
\begin{equation*}
   u_n:= I_0(f,g) - \la L_0(V u_{n-1}),\qquad n=0,1,2,...
\end{equation*}
Then we have the following 
\begin{Theorem} \label{Th:V}
With $f,g$ and $V$ as above for any $m>3$ and $k>2$ the sequence $u_n$ converges (in norm) in $\Linftx{p}$ for $p=\min(k,m-1)$ provided $\la<C_{p,k}^{-1}$. The limit $u:=\lim_{n\ra\infty} u_n$ satisfies
\begin{equation*}
  |u(t,x)| \leq \frac{C}{\norm{t+|x|}\norm{t-|x|}^{p-1}},\qquad \forall (t,x)\in\RTX
\end{equation*}
with  some positive constant $C$ depending only on $f_0, f_1, g_0, \la$ and $k,m$.
\end{Theorem}
This theorem was proved first for classical solutions in \cite{Strauss-T} and later generalized to weak solutions in \cite{NS-WaveDecay} and stated in a more detailed form, which will be important here. We cite the essential part of the proof because some of the presented estimates will be used later.
\begin{proof}
For $g,\nabla f\in\Linfx{m}$ and $f\in\Linfx{m-1}$ with $m>3$, from lemma \ref{Lem:init-data}, we get $u_0=I_0(f,g)\in\Linftx{m-1}$. Next, observe that if $u_n\in\Linftx{p}$ with some $p>1$ then
\begin{equation*}
  \|\<x\>^k V u_n\|_\Linftx{p} \leq \|\<x\>^k V\|_\Linf \|u_n\|_\Linftx{p} = \|u_n\|_\Linftx{p}<\infty
\end{equation*}
and from lemma \ref{Lem:source} with $F\equiv V u_n$ we get $L_0(V u_n)\in\Linftx{p}$ when $p\leq k$. Because $\Linftx{p_1}\subset\Linftx{p_2}$ when $p_1\geq p_2$, we get $u_{n+1}\in\Linftx{p}$ with $p\leq \min(m-1,k)$. By induction we obtain $u_n\in\Linftx{p}$ for every $n=0,1,2,...$ with the optimal value $p:=\min(m-1,k)$. Then, we have
\begin{equation*}
\begin{split}
  \|u_{n+1}-u_n\|_\Linftx{p} &= \la \|L_0(-V(u_n-u_{n-1}))\|_\Linftx{p} \leq
  \la C_{p,k} \|\<x\>^k V (u_n-u_{n-1})\|_\Linftx{p} \\ &\leq 
  \la C_{p,k} \|\<x\>^k V\|_\Linf \|u_n-u_{n-1}\|_\Linftx{p} \leq
  \la C_{p,k} \|u_n-u_{n-1}\|_\Linftx{p}
\end{split}
\end{equation*}
again making use of lemma \ref{Lem:source} with $\norm{x}^k F\equiv -\norm{x}^k V (u_n-u_{n-1}) \in \Linftx{p}$.
For $\delta:=\la\, C_{p,k}<1$ the iteration is a contraction in the normed space $\Linftx{p}$. A simple argument shows that the sequence $u_n$ is Cauchy. We have 
\begin{equation} \label{V:u_n-u_(n-1)}
  \|u_{n+1}-u_n\|_\Linftx{p} \leq \delta^{n+1} \|u_0-u_{-1}\|_\Linftx{p} = 
  \delta^{n+1} \|I_0(f,g)\|_\Linftx{p}
\end{equation}
and for $n'>n$
\begin{equation} \label{V:n-m}
\begin{split}
  \|u_{n'}-u_n\|_\Linftx{p} &\leq \sum_{j=0}^{n'-n-1} \|u_{n+j+1}-u_{n+j}\|_\Linftx{p} 
  \leq \sum_{j=0}^{n'-n-1} \delta^{j+n+1} \|I_0(f,g)\|_\Linftx{p}\\
  &\leq \frac{\delta^{n+1}}{1-\delta} \|I_0(f,g)\|_\Linftx{p}.
\end{split}
\end{equation}
This expression can be made arbitrarily small (smaller than any $\epsilon>0$) for all $n,m>M(\epsilon)$. Hence, $u_n$ is a Cauchy sequence in $\Linftx{p}$ which is Banach and $u_n$ has a limit $u\in\Linftx{p}$ satisfying
\begin{equation} \label{V:u-sol}
  u=I_0(f,g)-\la L_0(Vu).
\end{equation}
This equation is equivalent to the wave equation \refa{V:wave-eq} with the initial data \refa{fg-bound}. Finally, we find the $\Linftx{p}$-norm of $u$
\begin{equation*}
  \|u\|_\Linftx{p}\leq \|I_0(f,g)\|_\Linftx{p}+\la \|L_0(Vu)\|_\Linftx{p}
  \leq C_m(f_0+f_1+g_0)+\la\, C_{p,k} \|u\|_\Linftx{p},
\end{equation*}
thus
\begin{equation*}
  \|u\|_\Linftx{p}\leq \frac{C_m(f_0+f_1+g_0)}{1-\la\, C_{p,k}} \equiv C.
\end{equation*}
\end{proof}

\subsection{Perturbation series}

Now, we define a perturbation series by
\begin{equation*}
   u = \sum_{n=0}^\infty \la^n v_n
\end{equation*}
and insert into the wave eq. \refa{V:wave-eq}. It leads to the following perturbation scheme
\begin{alignat}{4} \label{pert-V-0}
  \Box v_0 &= 0,&\qquad (v_0,\dot{v}_0)(0)&=(f,g)&\qquad&\ra&\qquad v_0&=I_0(f,g) \\ \label{pert-V-n}
   \Box v_{n+1} &= -V v_n,&\qquad (v_{n+1},\dot{v}_{n+1})(0)&=(0,0)&\qquad&\ra&\qquad v_{n+1} &= -L_0(V v_n)
\end{alignat}
Due to linearity of \refa{V:wave-eq} it turns out that the partial sums
\begin{equation*}
   \sum_{k=0}^n \la^k v_k = u_n
\end{equation*}
give the elements $u_n$ obtained above by the iteration technique, so both methods (if they work) are equivalent. Theorem \ref{Th:V} implies convergence in $\Linftx{p}$ for $\la<C_{p,k}^{-1}$ with $k>2$, $m>3$ and $p:=\min(k,m-1)$.

From \refa{V:u_n-u_(n-1)} in the proof of Theorem \ref{Th:V} it follows that 
\begin{equation*}
  \| v_n\|_\Linftx{p} = \frac{\|u_{n}-u_{n-1}\|_\Linftx{p}}{\la^n} \leq (C_{p,k})^{n} \|I_0(f,g)\|_\Linftx{p},
\end{equation*}
hence $v_n \in \Linftx{p}$ for all $n\geq 0$. Observe, however, that in the case when $m-1>k=p$ we have at the lowest order $v_0=u_0$ a better decay estimate, namely $v_0\in\Linftx{m-1}$ (see first line of the proof). The reason that $v_0$ decays faster is that its decay comes only from initial data and is not influenced by the potential. At all higher orders, $v_n (n=1,2,...)$ contain the contribution from the scattering on the potential and are only in $\Linftx{k}$. Since $u\in\Linftx{p}=\Linftx{k}$, we expect that all $u_n$ starting from $u_1\in\Linftx{k}$ predict qualitatively correct asymptotic behaviour of $u$ while the lowest order $u_0\in\Linftx{m-1}$ fails in this. This becomes especially evident for initial data with compact support, for which $u_0\in\Linftx{q}$ with arbitrarily big $q$, but $u_1, u_2, ...\in\Linftx{k}\ni u$.


Knowing that the perturbation series converges for some $\la$ we can estimate the error of the $n$-th perturbation's order relative to the exact solution by estimating the sum of all higher order terms. For the convergent sequence $u_n$ we use the relation \refa{V:n-m} which holds also in the limit $n'\ra\infty$, $u_{n'}\ra u$ and gives
\begin{equation*} 
  \|u-u_n\|_\Linftx{p} \leq \frac{\delta^{n+1}}{1-\delta} \|I_0(f,g)\|_\Linftx{p}.
\end{equation*}
It provides a pointwise bound on the error
\begin{equation} \label{pointwise-error-bound}
  |u(t,x)-u_n(t,x)| \leq \frac{(C_{p,k}\la)^{n+1}}{1-C_{p,k}\la}\cdot 
  \frac{C_m\cdot(f_0+f_1+g_0)}{\norm{t+|x|}\norm{t-|x|}^{p-1}} \qquad \forall (t,x)\in\RTX.
\end{equation}

\subsection{Optimal decay estimate}

In this section we sketch a proof how, under some conditions, the optimal decay estimate and precise asymptotic behaviour of the solution $u$ can be deduced from the behaviour of low order perturbations.
This will be studied in more detail in a forthcoming publication \cite{NS-PB_Tails} dealing with spherical symmetry where the lowest perturbation orders can be calculated almost explicitly.

Consider first the case $m-1>k=p$, i.e. when the rate of decay of $u$ is dominated by scattering on the potential (and not by decay of the initial data). We have $u_0=v_0\in\Linftx{m-1}$ and $v_n\in\Linftx{p}$ for $n\geq 1$. 
Below, we show that if the asymptotic behaviour of $v_1$ is such as provided by its estimate (i.e. $p$ in the norm $\Linftx{p}$ is optimal) then theorem \ref{Th:V} gives an optimal estimate for $u\in\Linftx{p}$ with the same decay rate $p$.
Here, we consider only the asymptotics in direction of timelike infinity (the case of spatial infinity can be treated similarly). Assume, we are able to show (by some explicit calculation, like in \cite{NS-PB_Tails}) that $v_1(t,x) = L_0(V v_0) \cong c_1(x) t^{-p}\neq 0$ for $t\gg 1$, where $c_1(x)$ is independent on $\la$. The approximation sign means that for every small $\eta>0$ and every $x\in\R^3$ there is a $T_0(x,\eta)>0$ such that for all $t>T_0(x,\eta)$ the relative error is small, i.e.
\begin{equation} \label{def0-approx}
  \left|v_1(t,x) - \frac{c_1(x)}{t^p}\right|\leq \eta \frac{|c_1(x)|}{t^p}.
\end{equation}
From \refa{pointwise-error-bound} with $u_1=v_0+\la v_1$ we have
\begin{equation*}
  |u(t,x)-v_0(t,x)-\la v_1(t,x)| \leq \frac{(C_{p,k}\la)^{2}}{1-C_{p,k}\la}\cdot 
  \frac{C_m\cdot(f_0+f_1+g_0)}{\norm{t+|x|}\norm{t-|x|}^{p-1}}=:\Delta_1(t,x)
\end{equation*}
for all $(t,x)\in\RTX$. A simple inequality\footnote{It follows immediately from the Bernoulli's inequality.}
\begin{equation} \label{ineq-Bernoulli}
  \frac{1}{(1-\zeta)^\sigma} \leq \frac{1}{1-\sigma\zeta} = 1+\frac{\sigma\zeta}{1-\sigma\zeta} \leq 2,\qquad
  \forall \zeta \leq 1/(2\sigma),\quad \sigma>1
\end{equation}
implies
\begin{equation*}
  \frac{1}{\norm{t-|x|}^q} = \frac{1}{(1+t)^q \left(1-\frac{|x|}{1+t}\right)^q} \leq \frac{2}{(1+t)^q} 
\end{equation*}
for $\zeta:=|x|/(1+t)\leq 1/(2q)$, hence is true for all $t\geq 2q |x|$. The error term can be estimated
\begin{equation*}
  \Delta_1(t,x)
  \leq 2\,(C_{p,k} \la)^2 \frac{2\, C_m\cdot(f_0+f_1+g_0)}{(1+t)^p} \equiv
  \widetilde{C} \frac{\la^2}{(1+t)^p},
\end{equation*}
where we have twice used \refa{ineq-Bernoulli} for $t\geq 2(p-1) |x|$ and $\la \leq 1/(2\,C_{p,k})$.
Further,
\begin{equation*}
  \widetilde{C} \frac{\la^2}{(1+t)^p}  
  \leq \eta \la \frac{|c_1(x)|}{t^p} 
\end{equation*}
provided $\la$ is small enough such that $\la\leq \Lambda_\eta(x):=\eta c_1(x)/\widetilde{C}$.
Again from \refa{ineq-Bernoulli} we get
\begin{equation*}
  |v_0(t,x)|\leq \frac{c_0}{\norm{t+|x|} \norm{t-|x|}^{m-2}} \leq 
  \frac{2 c_0}{(1+t)^{m-1}} 
\end{equation*}
for all $t \geq 2(m-2)|x|$. Then,
\begin{equation*}
  |v_0(t,x)|\leq  \frac{2 c_0}{(1+t)^{m-1}} 
  \leq \eta \la \frac{|c_1(x)|}{t^{p}} 
\end{equation*}
provided $t>T_1(x,\eta,\la)$ is big enough, such that $t^{m-1-p} \geq 2 c_0/(\eta\, \la\, |c_1(x)|)$.

Finally, we arrive at the statement that for every small $\eta>0$ and every $x\in\R^3$, for sufficiently small $\la\leq \min[\Lambda_\eta(x), 1/(2\,C_{p,k})]$, and for sufficiently big $t>\max[T_0(x,\eta),T_1(x,\eta,\la),2(m-2)|x|]$ we have
\begin{equation*}
\begin{split}
  \left|u(t,x)-\la\frac{c_1(x)}{t^p}\right| &\leq 
  \left|u(t,x)-v_0(t,x)-\la v_1(t,x)\right| + |v_0(t,x)| + \la \left|v_1(t,x)-\frac{c_1(x)}{t^p}\right|\\
  &\leq  3 \eta\la \frac{|c_1(x)|}{t^p},
\end{split}
\end{equation*}
that is, for $p=k$,
\begin{equation*}
  u(t,x)\cong \la \frac{c_1(x)}{t^k}.
\end{equation*}
That gives a precise information about the time-decay of $u(t,x)$ and shows that the estimate in theorem \ref{Th:V} is optimal (for $t\gg |x|$).

In the case $p=m-1\leq k$ the decay rate of $u$ is determined by the decay of (long range) initial data and all $v_n\in\Linftx{p}$. Analogously, if we can show that $v_0(t,x) \cong c_0(x) t^{-p} \neq 0$ for $t\gg 1$ then we can bound all higher perturbation orders for sufficiently small $\la$ and big $t$ by the same expression multiplied by an arbitrarily small $\eta$. To this aim we use again \refa{pointwise-error-bound}
\begin{equation*}
  |u(t,x)-v_0(t,x)| \leq \frac{C_{p,k}\la}{1-C_{p,k}\la}\cdot 
  \frac{C_m\cdot(f_0+f_1+g_0)}{\norm{t+|x|}\norm{t-|x|}^{p-1}}=:\Delta_0(t,x) \quad \forall (t,x)\in\RTX,
\end{equation*}
and bound $\Delta_0(t,x)$ by $\eta\, |c_0(x)| t^{-p}$ as above. It leads to
\begin{equation*}
  \left|u(t,x)-\frac{c_0(x)}{t^p}\right| \leq 
  \left|u(t,x)-v_0(t,x)\right| + \left|v_0(t,x)-\frac{c_0(x)}{t^p}\right| 
  \leq  2 \eta \frac{|c_0(x)|}{t^p},
\end{equation*}
what for $p=m-1$ gives
\begin{equation*}
  u(t,x)\cong \frac{c_0(x)}{t^{m-1}}.
\end{equation*}
That again gives a precise information about the time-decay of $u(t,x)$ and shows that the estimate in theorem \ref{Th:V} is optimal (for $t\gg |x|$).


\section{Nonlinear case without the potential term}

Now, we consider a nonlinear wave equation of the form
\begin{equation} \label{Fu:wave-eq}
   \Box u = F(u)
\end{equation}
subject to initial data $(f,g)$ satisfying \refa{fg-bound} with $f_0, f_1, g_0 < \eps$. The nonlinear term obeys $|F(u)|\leq F_1 |u|^p$ for $|u|<1$ and $|F(u)-F(v)|\leq F_2 |u-v| \max(|u|,|v|)^{p-1}$. The second condition is satisfied e.g. for $F(u)=u^p$ with $F_2 = p$ or for $F\in\C^1$ such that $|F'(u)|\leq F_2 |u|^{p-1}$ for $|u|<1$.

\subsection{Iteration}
We define an iteration scheme 
\begin{equation*}
   u_{0} := 0,
\end{equation*}
\begin{equation} \label{Fu-iter}
   u_{n+1} := I_0(f,g) + L_0(F(u_n)), \qquad n\geq 0.
\end{equation}
For it we have the following 
\begin{Theorem} \label{Th:Fu}
With $f,g$ and $F(u)$ as above for any $m>3$, $p> 1+\sqrt{2}$ and sufficiently small $\eps$ the sequence $u_n$ converges (in norm) in $\Linftx{q}$ for $q=\min(p-1,m-1)$ to the solution $u$ of the equation \refa{Fu:wave-eq}. The limit $u:=\lim_{n\ra\infty} u_n$ satisfies
\begin{equation*}
  |u(t,x)| \leq \frac{C}{\norm{t+|x|}\norm{t-|x|}^{q-1}},\qquad \forall (t,x)\in\RTX
\end{equation*}
with  some positive constant $C$ depending only on $p,m$ and $\eps$.
\end{Theorem}

\begin{proof}
For $g,\nabla f\in\Linfx{m}$ and $f\in\Linfx{m-1}$ with $m>3$ from lemma \ref{Lem:init-data} we get $u_1=I_0(f,g)\in\Linftx{m-1}$. Next, if some $u_n\in\Linftx{q}$ with some $q>1$ then, since $L_0$ is a positive operator\footnote{In fact $L_0=\Box^{-1}$ is a measure on $\RTX$ and therefore has a positive kernel. Then, $L_0(F)\geq 0$ if $F\geq 0$.}, we have $|L_0(F(u_n))|\leq F_1 L_0(|u_n|^p)$  and from lemma \ref{Lem:power-p} we get
\begin{equation*}
  \|L_0(F(u_n))\|_\Linftx{q} \leq F_1 \|L_0(|u_n|^p)\|_\Linftx{q} \leq F_1 C \|u_n\|_\Linftx{q}^p,
\end{equation*}
and hence $L_0(F(u_n))\in\Linftx{q}$ when $q\leq p-1$. Then, $u_{n+1}\in\Linftx{m-1}\cap\Linftx{q}=\Linftx{q}$ for $q:=\min(m-1,p-1)$. Hence, by induction we obtain $u_n\in\Linftx{q}$ for every $n=0,1,2,...$ and 
\begin{equation*}
  \|u_{1}\|_\Linftx{q}\leq \|I_0(f,g)\|_\Linftx{q} \leq C_m (f_0+ f_1+g_0) \leq 3 C_m \eps
\end{equation*}
\begin{equation*}
  \|u_{n+1}\|_\Linftx{q}\leq \|I_0(f,g)\|_\Linftx{q} + \|L_0(F(u_n))\|_\Linftx{q}
  \leq 3 C_m \eps + F_1 C \|u_n\|_\Linftx{q}^p.
\end{equation*}
Choose $\eps>0$ such that $F_1 C (6 C_m)^p \eps^{p-1} < 3 C_m\cdot \min(1,2F_1/F_2)$. Then,
\begin{equation*}
  \|u_{1}\|_\Linftx{q}\leq 6 C_m \eps
\end{equation*}
\begin{equation*}
  \|u_{n}\|_\Linftx{q}\leq 6 C_m \eps \quad \Rightarrow \quad
  \|u_{n+1}\|_\Linftx{q}\leq 3 C_m \eps + F_1 C (6 C_m \eps)^p 
  \leq 6 C_m \eps,
\end{equation*}
hence $\|u_{n}\|_\Linftx{q}\leq 6C_m\eps$ for all $n\geq 1$. As next, we show convergence of the sequence $u_n$ by demonstrating that it is Cauchy.
\begin{equation*}
\begin{split}
  \|u_{n+1}-u_n\|_\Linftx{q} &= \|L_0(F(u_n)-F(u_{n-1}))\|_\Linftx{q}\\
  &\leq F_2 \|L_0(|u_n-u_{n-1}|\max(|u_n|,|u_{n-1}|)^{p-1})\|_\Linftx{q}\\
  &\leq F_2 C \||u_n-u_{n-1}|\max(|u_n|,|u_{n-1}|)^{p-1}\|_\Linftx{q}\\  
  &\leq F_2 C (6C_m\eps)^{p-1}\|u_n-u_{n-1}\|_\Linftx{q} 
  = \delta \|u_n-u_{n-1}\|_\Linftx{q},
\end{split}
\end{equation*}
with $\delta:=F_2 C (6C_m\eps)^{p-1}<1$, hence the iteration is a contraction in the normed space $\Linftx{q}$ and $u_n$ is a Cauchy sequence, because
\begin{equation} \label{Fu:u_n-u_(n-1)}
  \|u_{n+1}-u_n\|_\Linftx{q} \leq \delta^{n} \|u_1-u_{0}\|_\Linftx{q} = 
  \delta^{n} \|I_0(f,g)\|_\Linftx{q} \leq \delta^{n}\, 3 C_m \eps
\end{equation}
and for any $n'>n$
\begin{equation} \label{Fu:u_n-u_n}
  \|u_{n'}-u_n\|_\Linftx{q} \leq \sum_{j=0}^{n'-n-1} \|u_{n+j+1}-u_{n+j}\|_\Linftx{q} 
  \leq \sum_{j=0}^{n'-n-1} \delta^{j+n}\, 3C_m \eps 
  \leq \frac{\delta^{n}}{1-\delta}\, 3C_m \eps.
\end{equation}
Since $\Linftx{q}$ is Banach, $u_n$ has a limit $u\in\Linftx{q}$ satisfying
\begin{equation} \label{Fu:u-sol}
  u=I_0(f,g)+L_0(F(u))
\end{equation}
and solving the wave equation \refa{Fu:wave-eq} with the initial data \refa{fg-bound}. Its $\Linftx{q}$-norm satisfies
\begin{equation} \label{u-norm-Fu}
  \|u\|_\Linftx{q}\leq 6C_m\eps.
\end{equation}
\end{proof}
From \refa{Fu:u_n-u_n} it follows, in the limit $n'\ra\infty$ an error bound
\begin{equation} \label{Fu:u-u_n}
  \|u-u_n\|_\Linftx{q} \leq \frac{\delta^{n}}{1-\delta} \,3C_m\eps \leq C \eps^{(p-1)n+1}
\end{equation}
for small $\eps$.

\subsection{Perturbation series}

In order to be able to construct a well-defined perturbation scheme to all orders we have to assume additionally that $F(u)$ is analytic at $u=0$,
its Taylor series starts at power $p\geq 3$ and has convergence radius $R_F>0$. Then, for small initial data
\begin{equation} \label{Fu-pert-initdata}
  (u,\dot{u})(0) = (\eps f,\eps g)
\end{equation}
we introduce a perturbation series for representing the solution of \refa{Fu:wave-eq} 
\begin{equation} \label{Fu-pert-ser}
   u = \sum_{n=1}^\infty \eps^n v_n.
\end{equation}
After inserting it into \refa{Fu:wave-eq} and collecting terms according to powers of $\eps$ we obtain the following perturbation scheme
\begin{alignat}{4} \label{Fu-perteq1}
  \Box v_1 &= 0,&\qquad (v_1,\dot{v}_1)(0)&=(f,g)&\quad&\ra&\quad 
  v_1&=I_0(f,g) \\  \label{Fu-perteq}
  \Box v_{n+1} &= F_n(v_1,...,v_n),&\qquad (v_{n+1},\dot{v}_{n+1})(0)&=(0,0)&\quad&\ra&\quad
  v_{n+1} &= L_0(F_n(v_1,...,v_n)),
\end{alignat}
for $n\geq 1$, where $F_n$ result from collecting the nonlinear terms with the same powers of $\eps$
\begin{equation} \label{Fu-Fn}
  F_n(v_1,...,v_n)=
  \sum_k 
  a^n_k v_1^{\alpha^{n,1}_k} \cdots v_n^{\alpha^{n,n}_k},
\end{equation}
where $\alpha^{n,m}_k\in\N$ satisfy $\sum_{m=1}^n m\alpha^{n,m}_k = n+1$ and $\sum_{m=1}^n \alpha^{n,m}_k\geq p$ for every $n,k$.

We call this expansion a ``zero background'' case because the zero-order term $v_0$ is absent. If a $v_0$ term were present in the series above (i.e. the summation started at $n=0$), we would have an additional equation $\Box v_0 = F(v_0)$ which is truly nonlinear (opposite to the above system of linear wave equations with source terms). Its solution $v_0$ represents a ``background'' around which the perturbations $v_n$ are calculated.

Below we show that the perturbation series converges to the solution $u$ of the nonlinear wave equation \refa{Fu:wave-eq} and has a positive convergence radius.

\begin{Theorem} \label{Th:Fu-pert}
With $f,g$ and $F(u)$ as above for any $m>3$, $p> 1+\sqrt{2}$ and sufficiently small $\eps$ the series defined in \refa{Fu-pert-ser}-\refa{Fu-perteq} converges (in norm) in $\Linftx{q}$ for $q=\min(p-1,m-1)$ to the solution of the equation \refa{Fu:wave-eq} with initial data \refa{Fu-pert-initdata}.
\end{Theorem}

\begin{proof}
For $g,\nabla f\in\Linfx{m}$ and $f\in\Linfx{m-1}$ with $m>3$ from lemma \ref{Lem:init-data} we get $v_1= I_0(f,g)\in\Linftx{m-1}$ with
\begin{equation*}
   \|v_1\|_\Linftx{m-1} \leq C_m(f_0+f_1+g_0) =: D < \infty.
\end{equation*}
Next, we prove by induction $\Linftx{q}$ bounds for all $n\geq 1$ with some $q>1$. Assume that for a given $n\geq 1$ we have $v_m\in\Linftx{q}$ for all $m\leq n$. Then, using \refa{Fu-Fn}, we get
\begin{equation*}
  \|v_{n+1}\|_\Linftx{q} = \|L_0(F_n(v_1,...,v_n))\|_\Linftx{q} 
  \leq \sum_k |a^n_k|\cdot \|L_0(|v_1^{\alpha^{n,1}_k} \cdots v_n^{\alpha^{n,n}_k}|)\|_\Linftx{q}.  
\end{equation*}
Observe, that $|v_1^{\alpha^{n,1}_k} \cdots v_n^{\alpha^{n,n}_k}|=\left(\sqrt[p]{|v_1^{\alpha^{n,1}_k} \cdots v_n^{\alpha^{n,n}_k}|}\right)^p$ and $\sqrt[p]{|v_1^{\alpha^{n,1}_k} \cdots v_n^{\alpha^{n,n}_k}|}\in\Linftx{q}$, because of the following estimate for $b_1+\ldots+b_n=B\geq 1$ and $b_m\geq 0$, $m=1,...,n$
\begin{equation*}
\begin{split}
  &\|w_1^{b_1} \cdots w_n^{b_n}\|_\Linftx{q} 
  = \|\norm{t+|x|}\norm{t-|x|}^{q-1} w_1^{b_1} \cdots w_n^{b_n}\|_\Linf \\
  \leq& \|\norm{t+|x|}^{b_1/B}\norm{t-|x|}^{(q-1)b_1/B} w_1^{b_1}\|_\Linf \cdots 
        \|\norm{t+|x|}^{b_n/B}\norm{t-|x|}^{(q-1)b_n/B} w_n^{b_n}\|_\Linf \\
  =&    \|\norm{t+|x|}^{1/B}\norm{t-|x|}^{(q-1)/B} w_1\|_\Linf^{b_1} \cdots 
        \|\norm{t+|x|}^{1/B}\norm{t-|x|}^{(q-1)/B} w_n\|_\Linf^{b_n} \\
  \leq& \|\norm{t+|x|}\norm{t-|x|}^{(q-1)} w_1\|_\Linf^{b_1} \cdots 
        \|\norm{t+|x|}\norm{t-|x|}^{(q-1)} w_n\|_\Linf^{b_n} \\
  =&    \|w_1\|_\Linftx{q}^{b_1} \cdots \|w_n\|_\Linftx{q}^{b_n},
\end{split}
\end{equation*}
which used for $b_m:=\alpha^{n,m}_k/p$ with
$B:=\alpha^{n,1}_k/p+\ldots+\alpha^{n,n}_k/p\geq 1$ gives
\begin{equation*}
\begin{split}
  \left\|\sqrt[p]{|v_1^{\alpha^{n,1}_k} \cdots v_n^{\alpha^{n,n}_k}|}\right\|_\Linftx{q} &=
  \left\||v_1|^{\alpha^{n,1}_k/p} \cdots |v_n|^{\alpha^{n,n}_k/p}\right\|_\Linftx{q} \leq
  \|v_1\|_\Linftx{q}^{\alpha^{n,1}_k/p} \cdots \|v_n\|_\Linftx{q}^{\alpha^{n,n}_k/p} 
  < \infty.
\end{split}
\end{equation*}
Then, for $q\leq p-1$ we can use lemma \ref{Lem:power-p} with $u:=\sqrt[p]{|v_1^{\alpha^{n,1}_k} \cdots v_n^{\alpha^{n,n}_k}|}$ to obtain
\begin{equation*}
\begin{split}
  \|v_{n+1}\|_\Linftx{q} &
  \leq C\sum_k |a^n_k|\cdot \left\|\sqrt[p]{|v_1^{\alpha^{n,1}_k} \cdots v_n^{\alpha^{n,n}_k}|}\right\|_\Linftx{q}^p 
  = C\sum_k |a^n_k|\cdot \left\|v_1^{\alpha^{n,1}_k} \cdots v_n^{\alpha^{n,n}_k}\right\|_\Linftx{q} \\
  &\leq C \sum_k |a^n_k|\cdot \|v_1\|_\Linftx{q}^{\alpha^{n,1}_k} \cdots \|v_n\|_\Linftx{q}^{\alpha^{n,n}_k}.
\end{split}
\end{equation*}
Unfortunately, we were not able to find an estimate for $\sum_k |a^n_k|$ being good enough to prove a geometric growth of $\|v_{n}\|_\Linftx{q}$ and guaranteeing convergence of the series \refa{Fu-pert-ser}. If one tries so, e.g. assuming $\|v_m\|_\Linftx{q}\leq D^m$ for all $m\leq n$, then
\begin{equation*}
\begin{split}
  \|v_{n+1}\|_\Linftx{q} 
  &\leq C \sum_k |a^n_k|\cdot \|v_1\|_\Linftx{q}^{\alpha^{n,1}_k} \cdots \|v_n\|_\Linftx{q}^{\alpha^{n,n}_k} 
  \leq C \sum_k |a^n_k|\cdot D^{(1\alpha^{n,1}_k+\ldots+n\alpha^{n,n}_k)} \\
  &= C D^{n+1} \sum_k |a^n_k|.
\end{split}
\end{equation*}
The best estimate we were able to find is $\sum_k |a^n_k|\leq \widetilde{C} n^p$ (imposing further assumptions on $F(u)$) which does not allow to close the induction argument. Therefore, we choose a different way and use some trick, relating the wave equation to an algebraic one.

To this goal, we need to relate the coefficients of the power series for $F(u)$
\begin{equation*}
  F(u)=\sum_{n=p}^\infty b_n u^n,
\end{equation*}
which converges for $|u|<R_F$, to the expansion coefficients $a^n_k$ which result from a formal insertion of the series $u=\sum_{k=1}^\infty \eps^k v_k$ into $F(u)$
\begin{equation} \label{Fu-trick-F(u)-ser}
  F\left(\sum_{k=1}^\infty \eps^k v_k\right) \equiv 
  \sum_{n=p-1}^\infty \eps^{n+1} F_n(v_1,...,v_n) \equiv
  \sum_{n=p-1}^\infty \eps^{n+1} \sum_k a^n_k\; v_1^{\alpha^{n,1}_k} \cdots v_n^{\alpha^{n,n}_k}.
\end{equation}
By some manipulation of sums 
we obtain
\begin{equation*}
  a^n_k = b_{\alpha^{n,1}_k+\ldots+\alpha^{n,n}_k} 
  \begin{pmatrix} \alpha^{n,1}_k+\ldots+\alpha^{n,n}_k \\ \alpha^{n,1}_k,\ldots,\alpha^{n,n}_k
  \end{pmatrix},    
\end{equation*}
where the symbol in delimiters represents the multinomial coefficient. Since there is an analogous relation between the absolute values of the coefficients
\begin{equation*}
  |a^n_k| = |b_{\alpha^{n,1}_k+\ldots+\alpha^{n,n}_k}| 
  \begin{pmatrix} \alpha^{n,1}_k+\ldots+\alpha^{n,n}_k \\ \alpha^{n,1}_k,\ldots,\alpha^{n,n}_k
  \end{pmatrix}
\end{equation*}
we observe that the series \refa{Fu-trick-F(u)-ser} with $a^n_k$ replaced by $|a^n_k|$ gives rise to a new function $\w{F}$
\begin{equation*}
  \sum_{n=p-1}^\infty \eps^{n+1} \sum_k |a^n_k|\; v_1^{\alpha^{n,1}_k} \cdots v_n^{\alpha^{n,n}_k}
  = \w{F}\left(\sum_{k=1}^\infty \eps^k v_k\right)
\end{equation*}
such that
\begin{equation} \label{series-F_}
  \w{F}(u)=\sum_{n=p}^\infty |b_n| u^n.
\end{equation}
$\w{F}(u)$ is also analytic at $u=0$ and the convergence radius is the same as that of $F(u)$, i.e. $R_{\w{F}}=R_F$ what follows from standard theory of analytic functions.

Now, instead of the system of estimates
\begin{align}
   \|v_1\|_\Linftx{q} &\leq  D, \\
  \|v_{n+1}\|_\Linftx{q} &\leq C \sum_k |a^n_k|\cdot \|v_1\|_\Linftx{q}^{\alpha^{n,1}_k} \cdots \|v_n\|_\Linftx{q}^{\alpha^{n,n}_k}
\end{align}
with $q:=\min(m-1,p-1)$, we consider a system of equations
\begin{align}
   w_1 &=  D, \label{eq-w1}\\
   w_{n+1} & = C \sum_k |a^n_k|\cdot w_1^{\alpha^{n,1}_k} \cdots w_n^{\alpha^{n,n}_k} \label{eq-w_n}
\end{align}
and it is easy to see (e.g. by induction) that $\|v_n\|_\Linftx{q}\leq w_n$ for all $n\geq 1$. Now comes the trick. Using the above relations we can find that this system is equivalent to
\begin{equation} \label{algebraic-Fu-series}
  \sum_{n=1}^\infty \eps^n w_n = C \w{F}\left(\sum_{n=1}^\infty \eps^n w_n\right) + D \eps.
\end{equation}
Introducing $W=\sum_{n=1}^\infty \eps^n w_n$ we can write
\begin{equation} \label{algebraic-Fu}
  {W = C \w{F}(W) + D \eps}.
\end{equation}
Since $\w{F}(W)$ is analytic at $W=0$, so is $G(W)$
\begin{equation*}
  G(W):=\frac{W-C\w{F}}{D} = \eps
\end{equation*}
and also its inverse $G^{-1}(\eps)$ at $\eps=0$, because $G'(0)=1/D>0$ (see e.g. (real) analytic inverse function theorem in \cite{Hille}), what follows from the fact that the Taylor series for $\w{F}$ starts (as that for $F$) at the power at least $p>2$.
Then $G^{-1}(\eps)$ has a Taylor series with a positive convergence radius $R_{G^{-1}}>0$.
The solution $W(\eps)$ of \refa{algebraic-Fu} can be then represented by a convergent series for $|\eps|<R_{G^{-1}}$
\begin{equation} \label{ser-W-eps}
  W(\eps)=G^{-1}(\eps) = \sum_{n=1}^\infty \eps^n w_n.
\end{equation}
In order to guarantee that this series can act as a good argument of $\w{F}$ we choose
a possibly smaller radius $\w{R}\leq R_{G^{-1}}$ such that $|W(\eps)|<R_F$ for all $|\eps|<\w{R}$.
Then $\w{F}(W(\eps))$ can be represented by a convergent series \refa{series-F_} in $W(\eps)$. Finally, this allows us to insert this series into \refa{algebraic-Fu} and obtain first \refa{algebraic-Fu-series} and then the system \refa{eq-w1}-\refa{eq-w_n}.

Essential for the trick is that the series in \refa{ser-W-eps} converges for all $|\eps|<\w{R}$. Now, since $\|v_n\|_\Linftx{q}\leq w_n$ for all $n\geq 1$, we get from the comparison criterion that the series $\sum_{n=1}^\infty \eps^n \|v_{n}\|_\Linftx{q}$ converges as well for all $|\eps|<\w{R}$. 
Thus, the series \refa{Fu-pert-ser} converges in norm in $\Linftx{q}$ for all $|\eps|<\w{R}$ to some $\w{u}\in\Linftx{q}$ which satisfies
\begin{equation*}
\begin{split}
  \widetilde{u} &:= \sum_{n=1}^\infty \eps^{n} v_{n}
  = \eps I_0(f,g) + \sum_{n=1}^\infty \eps^{n+1} L_0(F_n(v_1,...,v_n)) \\
  &= \eps I_0(f,g) + L_0\left(F\left(\sum_{n=1}^\infty \eps^{n} v_n\right)\right) \\
  &= I_0(\eps f,\eps g) + L_0(F(\widetilde{u}))
\end{split}
\end{equation*}
what is equivalent to the wave equation \refa{Fu:wave-eq} with initial data \refa{Fu-pert-initdata}. Uniqueness of solutions follows easily from theorem \ref{Th:Fu}.

An important consequence of the convergence of $\sum_{n=1}^\infty \eps^n \|v_{n}\|_\Linftx{q}$ is that there exist constants $0<M<\infty$ and $\w{R}^{-1}<\rho<\infty$ such that $\|v_{n}\|_\Linftx{q}<M \rho^n$ for every $n\geq 1$.
\end{proof}

Since the introduction of the auxiliary parameter $\eps$ in the series expansion is only a way to generate the system of linear equations equivalent to the original nonlinear equation, we can now remove the parameter $\eps$ and replace the condition on the initial data by requiring $f_0, f_1, g_0 < \w{R}$. If we solve the system \refa{Fu-perteq1}-\refa{Fu-perteq} then we obtain a solution of the nonlinear wave equation \refa{Fu:wave-eq} by summing up the convergent series $\sum_{n=1}^\infty v_n = u$.

\subsection{Optimal decay estimate}

In the nonlinear case, the iteration sequence $u_n$ is different than the perturbation sequence $\widetilde{u}_n:=\sum_{m=1}^n v_n$, therefore the question whether information about the decay rate of $u$ can be read-off from the low order terms must be studied separately for both cases. On the one hand, in the iterative scheme the form of the source terms $F(u_n)$ is much simpler than that in the perturbative scheme, $F_n(v_1,...,v_n)$. On the other hand, in practise, it is much easier to calculate $v_n$'s than $u_n$'s.
Below, we address both situations.

Analogously like for the linear equation, we will have two cases depending on whether $m$ is smaller or bigger than $p$. In the first case, the initial data will dominate the late-time decay rate of $u$, in the second case the power $p$ of the nonlinearity, through nonlinear scattering, will determine the the decay rate of $u$. 

\subsubsection{Iteration}

In analogy to the linear case, basing on a decay information for some low order term in the iteration sequence and on its error bound we find the exact decay rate of $u$. From the error bound \refa{Fu:u-u_n} it follows for large $t$
\begin{equation*}
  |u(t,x)-u_n(t,x)|\leq \frac{C\eps^{(p-1)n+1}}{\norm{t+|x|}\norm{t-|x|}^{q-1}}
  \leq \frac{c_n(x)\eps^{(p-1)n+1}}{(1+t)^q}
\end{equation*}
where $q:=\min(p-1,m-1)$. 
If we are able to show that some $u_n(t,x)\cong \eps d_n(x) t^{-q}\neq 0$ for large $t$ (the asymptotic approximation is to be understood in the following sense:
\begin{equation} \label{def-approx}
  \exists_{\eta_0>0} \forall_{0<\eta<\eta_0} \exists_{T<\infty} \forall_{t>T} 
  \left| u_n(t,x) - \frac{\eps d_n(x)}{t^q}\right| < \eta \frac{\eps d_n(x)}{t^q},
\end{equation}
i.e. the relative error $\eta$ becomes arbitrarily small for sufficiently big $t$, cf. \refa{def0-approx}), then already $u_n$ shows the correct decay rate, identical with this of $u$, because then, choosing $\eta:=\eps^{(p-1)n}$, we get
\begin{equation*}
  \left|u(t,x)-\eps \frac{d_n(x)}{t^q}\right| \leq 
  \left|u(t,x)-u_n(t,x)\right| + \left|u_n(t,x)-\eps \frac{d_n(x)}{t^q}\right| \leq
  \eta \eps \frac{c_n(x)}{t^q} + \eta \eps \frac{d_n(x)}{t^q}
\end{equation*}
for sufficiently small $\eps$. Hence the decay rate of $u$ at late times is exactly $t^{-q}$.

In case when $m>p=q+1$, we have $u_1=I_0(f,g)\in\Linftx{m-1}$ and 
\begin{equation*}
  |u_1(t,x)|\leq \frac{C_m\cdot(f_0+f_1+g_0)}{\norm{t+|x|}\norm{t-|x|}^{m-2}} \leq
  \frac{2 C_m\cdot(f_0+f_1+g_0)}{(1+t)^{m-1}}
\end{equation*}
for $t\geq 2(m-2)|x|$,
hence it decays faster than $u$ and it cannot be shown that $u_1(t,x)\cong \eps d_1(x) t^{-q}$. It is expected that it will be true for $u_2\cong \eps d_2(x) t^{-q}$, what means, that already $u_2$ would have the same rate of decay as $u$ (see \cite{NS-PB_Tails} for such results in spherical symmetry).

In case when $m \leq  p$, we have $q:=m-1$. Then it should be possible to show $u_1(t,x)\cong c_1(x) t^{-(m-1)}$, what means, that already $u_1$ would have the same rate of decay as $u$.

\subsubsection{Perturbation series}

The perturbation scheme \refa{Fu-perteq1}-\refa{Fu-perteq} can be written as
\begin{align}
  v_1&=I_0(f,g) \\
  v_2&=v_3=...=v_{p-1}=0\\
  v_{p} &= L_0(F_{p-1}(v_1,...,v_{p-1})) = a_0 L_0((v_1)^p)\\
  v_{n+1} &= L_0(F_n(v_1,...,v_n)),\qquad n\geq p.
\end{align}
Assume we are in the more interesting case $m>p$ where the tail results from the nonlinear scattering. Then, $v_1=I_0(f,g)\in\Linftx{m-1}$ and $v_n\in\Linftx{p-1}$ for $n\geq 2$. If we can show that $a_0 L_0((I(f,g))^p)\cong d_p(x) t^{-(p-1)}$, then already $v_p$ has the correct decay rate, identical with this of $u$. To prove it, we need to show that $\eps I_0(f,g)$ and $\eps^{n+1} L_0(F_n(v_1,...,v_n))$ for $n\geq p$ are small relative to $\eps^p d_p(x) t^{-(p-1)}$. 

Again, for $v_1=I_0(f,g)\in\Linftx{m-1}$ the situation is obvious, $\eps |v_1| = \eps |I_0(f,g)(t,x)|\leq c_1(x)\eps (1+t)^{-(m-1)}$ and it is much smaller than $\eps^p d_p(x) t^{-(p-1)}$ for sufficiently large $t$.

From the convergence proof for the perturbation series we know that there exist constants $M,\rho>0$ such that $\|v_{n}\|_\Linftx{p-1}\leq M \rho^n$ for all $n\geq 1$. Hence, we can estimate the remainder of the perturbation series
\begin{equation*}
  \left\|\sum_{m=p+1}^\infty \eps^m v_m\right\|_\Linftx{p-1} 
  \leq M \sum_{m=p+1}^\infty \eps^m \rho^m 
  \leq \frac{M \eps^{p+1} \rho^{p+1}}{1-\eps\rho} \leq C \eps^{p+1},
\end{equation*}
for sufficiently small $\eps<1/\rho$. It means that
\begin{equation*}
  \left|\sum_{m=p+1}^\infty \eps^m v_m(t,x)\right| \leq \frac{C \eps^{p+1}}{\norm{t+|x|}\norm{t-|x|}^{p-2}}
  \leq \frac{c(x) \eps^{p+1}}{(1+t)^{p-1}}
\end{equation*}
for big $t$ (and fixed $x$). Then
\begin{equation*}
  \left| u(t,x) -  \eps^p v_p(t,x)\right| \leq
  |\eps v_1(t,x)| + \left|\sum_{m=p+1}^\infty \eps^m v_m\right| \leq
  \frac{c_1(x) \eps}{t^{m-1}} + \frac{c(x) \eps^{p+1}}{t^{p-1}} 
\end{equation*}
hence with $v_p\cong d_p(x) t^{-(p-1)}$ and the relative error $\eta:=\eps$ (cf. \refa{def-approx} for the definition of ``$\cong$'')
\begin{equation*}
  \left|u(t,x) - \frac{d_p(x) \eps^p}{t^{p-1}}\right| \leq  
  \frac{c_1(x) \eps}{t^{m-1}} + \frac{[c(x) + d_p(x)] \eps^{p+1}}{t^{p-1}}.
\end{equation*}
For small $\eps$ and big $t$ such that $t^{m-p} \geq \frac{c_1(x)}{c(x)} \eps^{-p}$
it follows 
\begin{equation*}
  u(t,x) \cong \frac{d_p(x) \eps^p}{t^{p-1}}.
\end{equation*}
Thus $v_p$ dominates the perturbation series for large times and small $\eps$ and has the same decay rate as the full solution of the nonlinear wave equation $u$.

\section{Nonlinear case with the potential term}

Finally, let's consider a nonlinear wave equation with potential
\begin{equation} \label{Fu-V:wave-eq}
   \Box u + V u = F(u)
\end{equation}
subject to initial data $(f,g)$ satisfying \refa{fg-bound} with $f_0, f_1, g_0 < \eps$. The nonlinear term $F(u)$ is like in the previous section. 

\subsection{Iteration}
\subsubsection{Perturbative treatment of $V$}

As in the previous sections, we define an iteration
\begin{equation*}
   u_{0} := 0
\end{equation*}
\begin{equation*}
   u_{n+1} := I_0(f,g) - \la L_0(V u_n) + L_0(F(u_n)),\qquad n\geq 0
\end{equation*}
We have the following
\begin{Theorem} \label{Th:V-Fu}
With $f,g$, $V$ and $F(u)$ as above for any $m>3$, $k>2$, $p> 1+\sqrt{2}$, $\la<C_{q,k}^{-1}$ and sufficiently small $\eps$ the sequence $u_n$ converges (in norm) in $\Linftx{q}$ for $q=\min(p-1,k,m-1)$ to the solution $u$ of the equation \refa{Fu-V:wave-eq}. The limit $u:=\lim_{n\ra\infty} u_n$ satisfies
\begin{equation*}
  |u(t,x)| \leq \frac{C}{\norm{t+|x|}\norm{t-|x|}^{q-1}},\qquad \forall (t,x)\in\RTX
\end{equation*}
with  some positive constant $C$ depending only on $p,k,m,\la$ and $\eps$.
\end{Theorem}
The proof is a combination of proofs of theorems \ref{Th:V} and \ref{Th:Fu}, therefore we concentrate only on the points that differ.
\begin{proof}
For $g,\nabla f\in\Linfx{m}$ and $f\in\Linfx{m-1}$ with $m>3$ from lemma \ref{Lem:init-data} we get $u_1=I_0(f,g)\in\Linftx{m-1}$. Next, for $\delta:=\la C_{q,k}<1$ there exists $M>3/(1-\delta)>0$ and if $u_n\in\Linftx{q}$ with $\|u_n\|_\Linftx{q}\leq M C_m \eps$ for some $n\geq 1$ and $q>1$ then from lemmas \ref{Lem:init-data}-\ref{Lem:power-p} we get
\begin{equation*}
\begin{split}
  \|u_{n+1}\|_\Linftx{q} &\leq \|I_0(f,g)\|_\Linftx{q} + \la \|V u_n\|_\Linftx{q} +
  \|L_0(F(u_n))\|_\Linftx{q} \\
  &\leq C_m (f_0+ f_1+g_0) + \la C_{q,k} \|u_n\|_\Linftx{q} + F_1 C \|u_n\|_\Linftx{q}^p \\
  &\leq 3 C_m \eps + \delta M C_m \eps + F_1 C (M C_m)^p \eps^p < \infty
\end{split}
\end{equation*}
and hence $u_{n+1}\in\Linftx{q}$ if $q:=\min(m-1, k, p-1)$. By induction we obtain $u_n\in\Linftx{q}$ for every $n\geq 1$. For $\eps>0$ such that $F_1 C (M C_m)^p \eps^{p-1} \leq \min[(M(1-\delta)-3) C_m, M \delta(1-\delta) C_m F_1/F_2]$ we have
\begin{equation*}
  \|u_{1}\|_\Linftx{q}\leq 3 C_m \eps \leq M C_m \eps
\end{equation*}
\begin{equation*}
  \|u_{n+1}\|_\Linftx{q}\leq 3 C_m \eps  + \delta  M C_m \eps + (M(1-\delta)-3) C_m \eps
  = M C_m \eps,
\end{equation*}
hence $\|u_{n}\|_\Linftx{q}\leq M C_m\eps$ for all $n\geq 1$. Analogously like in the previous proofs, we arrive at
\begin{equation*}
\begin{split}
  \|u_{n+1}-u_n\|_\Linftx{q} &\leq  \|L_0(V(u_n-u_{n-1}))\|_\Linftx{q} +
   \|L_0(F(u_n)-F(u_{n-1}))\|_\Linftx{q}\\
  & \leq \la C_{q,k} \|u_n-u_{n-1}\|_\Linftx{q} + F_2 C (M C_m\eps)^{p-1}\|u_n-u_{n-1}\|_\Linftx{q} \\
  & \leq \delta \|u_n-u_{n-1}\|_\Linftx{q} + \delta(1-\delta) \|u_n-u_{n-1}\|_\Linftx{q} 
  = \delta' \|u_n-u_{n-1}\|_\Linftx{q},
\end{split}
\end{equation*}
where $\delta':=2\delta-\delta^2<1$. It follows that $u_n$ is a Cauchy sequence (see the above proofs) in the Banach space $\Linftx{q}$ and hence $u_n$ has a limit $u\in\Linftx{q}$ satisfying
\begin{equation} \label{Fu-V:u-sol}
  u=I_0(f,g)+L_0(Vu_n)+L_0(F(u))
\end{equation}
and solving the wave equation \refa{Fu-V:wave-eq} with the initial data \refa{fg-bound}. Its $\Linftx{q}$-norm satisfies
\begin{equation} \label{Fu-V:u-norm}
  \|u\|_\Linftx{q}\leq M C_m\eps
\end{equation}
with some (finite) constant $M>3/(1-\delta)>0$.
\end{proof}

Moreover, by analogous considerations like in the proof of theorem \ref{Th:Fu}, we find for $n'>n$
\begin{equation*}
  \|u_{n'}-u_n\|_\Linftx{q} \leq \frac{\delta'^n}{1-\delta'} 3 C_m \eps,
\end{equation*}
and in the limit $n'\ra\infty$
\begin{equation} \label{Fu-V:u-u_n}
  \|u-u_n\|_\Linftx{q} \leq \frac{\delta'^n}{1-\delta'} 3 C_m \eps.
\end{equation}

\subsubsection{Non-perturbative treatment of $V$}

Building on the above results we can also define an alternative iteration scheme
\begin{equation*}
   u_{0} := 0
\end{equation*}
\begin{equation*}
   u_{n+1} := I_V(f,g) + L_V(F(u_n)),\qquad n\geq 0
\end{equation*}
which is based on inversion of the operator $\Box+\la V$. According to the discussion in the introduction, it is equivalent to
\begin{equation*}
   u_{n+1} = I_0(f,g) + L_0(F(u_n)) - \la L_0(V u_{n+1}),\qquad n\geq 0.
\end{equation*}
It converges under the same conditions as in theorem \ref{Th:V-Fu}. The proof has the only difference that now we have
\begin{equation*}
\begin{split}
  \|u_{n+1}\|_\Linftx{q} &\leq \|I_0(f,g)\|_\Linftx{q} + \|L_0(F(u_n))\|_\Linftx{q} 
  + \la \|V u_{n+1}\|_\Linftx{q}\\
  &\leq C_m (f_0+ f_1+g_0) + F_1 C \|u_n\|_\Linftx{q}^p + \la C_{q,k} \|u_{n+1}\|_\Linftx{q}\\
  &\leq (1-\delta) M C_m \eps  + \delta \|u_{n+1}\|_\Linftx{q}
\end{split}
\end{equation*}
what gives
\begin{equation*}
  \|u_{n+1}\|_\Linftx{q} \leq \frac{(1-\delta) M C_m \eps}{1-\delta} = M C_m \eps.
\end{equation*}

\subsection{Perturbation series}

Definig a perturbation scheme for the nonlinear wave equation with potential \refa{Fu-V:wave-eq} 
\begin{equation} \label{Fu-V:pert-ser}
   u = \sum_{n=1}^\infty \eps^n v_n
\end{equation}
one encounters the problem of two scales which are introduced by parameters $\la$ measuring the strength of the potential and $\eps$ measuring the strength of the initial data. Therefore, we propose two ways of looking at the problem: in first, we treat the potential non-perturbatively, in second, we assign to $\la$ a scale of some power of $\eps$.

\subsubsection{Non-perturbative treatment of $V$ ($\la \sim \eps^0$)}

In this perturbation scheme we invert the operator $\Box+\la V$, thus treating $V$ in a non-perturbative way. For the sequence $v_n$ defined by
\begin{align} 
  v_1 &:=I_V(f,g) = I_0(f,g) - L_0(Vv_1) \label{V-Fu-1perteq1} \\
  v_{n+1} &:= L_V(F_n(v_1,...,v_n)) = L_0(F_n(v_1,...,v_n)) - L_0(V v_{n+1}),\qquad 
  n\geq 1 \label{V-Fu-1perteq}
\end{align}
we have the following
\begin{Theorem} \label{Th:V-Fu-pert1}
With $f,g, V$ and $F(u)$ as above for any $m>3$, $k>2$, $p> 1+\sqrt{2}$, $\la<C_{q,k}^{-1}$ and sufficiently small $\eps$ the series defined in \refa{Fu-V:pert-ser}-\refa{V-Fu-1perteq} converges (in norm) in $\Linftx{q}$ for $q=\min(p-1,k,m-1)$ to the solution of the equation \refa{Fu-V:wave-eq} with initial data \refa{Fu-pert-initdata}.
\end{Theorem}
\begin{proof}
The proof is essentially identical with this of theorem \ref{Th:Fu-pert} with the following differences. For $q:=\min(m-1,k,p-1)$ we obtain
\begin{equation*}
   \|v_1\|_\Linftx{m-1} \leq D + \delta \|v_1\|_\Linftx{m-1}, 
\end{equation*}
where $\delta:=\la C_{q,k}$ and hence
\begin{equation*}
   \|v_1\|_\Linftx{q} \leq \|v_1\|_\Linftx{m-1} \leq \frac{D}{1-\delta} < \infty.
\end{equation*}
The same modification regards all other inequalities
\begin{equation*}
  \|v_{n+1}\|_\Linftx{q} \leq C \sum_k |a^n_k|\cdot \|v_1\|_\Linftx{q}^{\alpha^{n,1}_k} \cdots \|v_n\|_\Linftx{q}^{\alpha^{n,n}_k} + \delta \|v_{n+1}\|_\Linftx{q}
\end{equation*}
which leads to
\begin{equation*}
  \|v_{n+1}\|_\Linftx{q} \leq \frac{C}{1-\delta} \sum_k |a^n_k|\cdot \|v_1\|_\Linftx{q}^{\alpha^{n,1}_k} \cdots \|v_n\|_\Linftx{q}^{\alpha^{n,n}_k}.
\end{equation*}
Repeating the trick used in the proof of theorem \ref{Th:Fu-pert}, we can relate this problem to the algebraic equation, which now becomes
\begin{equation} \label{algebraic-Fu-V}
  {W = \frac{C}{1-\delta}\; \w{F}(W) + \frac{D}{1-\delta}\; \eps}.
\end{equation}
Since $G(W)$ given by 
\begin{equation*}
  G(W):=\frac{(1-\delta) W-C\w{F}}{D} = \eps
\end{equation*}
is again analytic, so is $W(\eps)=G^{-1}(\eps)$, because $G'(0)=(1-\delta)/D>0$. Repeating the reasoning, we arrive at the conclusion that $\sum_{n=1}^\infty \eps^n \|v_n\|_\Linftx{q}$ has a positive radius of convergence.
It follows that the series \refa{Fu-V:pert-ser} converges in norm in $\Linftx{q}$ for all $\eps<\w{R}$ to the solution of \refa{Fu-V:wave-eq} with initial data \refa{Fu-pert-initdata}. 
Uniqueness follows easily from theorem \ref{Th:V-Fu}.
\end{proof}

We can, again, remove the auxiliary parameter $\eps$ and replace the condition on the initial data by $f_0, f_1, g_0 < \w{R}$. Then, the series $\sum_{n=1}^\infty v_n$ defined by \refa{V-Fu-1perteq1}-\refa{V-Fu-1perteq} converges to the solution of the nonlinear wave equation \refa{Fu-V:wave-eq}.

\subsubsection{Perturbative treatment of $V$ ($\la \sim \eps^a$)}

\def\law{\widetilde{\lambda}}

If we assume that the small scale of the potential's strength $\la$ is related to the small scale of the initial data, say $\la = \eps^a \law$ with $a\in\N_+$, then the power series Ansatz
\begin{equation} \label{Fu-V:pert-ser'}
   u = \sum_{n=1}^\infty \eps^n v_n
\end{equation}
inserted into the wave equation \refa{Fu-V:wave-eq} gives
\begin{align}
  v_{-n} &:= 0,\qquad n\geq 0 \label{V-Fu-2perteq-} \\
  v_1 &:=I_0(f,g) \label{V-Fu-2perteq1} \\
   v_{n+1} &:= -\law L_0(V v_{n+1-a}) + L_0(F_n(v_1,...,v_n)) \label{V-Fu-2perteq},\qquad n\geq 1.
\end{align}
This system is much more appropriate for numerical techniques, because the equation on $v_{n+1}$ is explicit, in contrast to the previous scheme, which includes implicit equations for $v_{n+1}$ (i.e. appearing on both sides). Moreover, if we choose $a:=p-1$, then the lowest nontrivial order, $v_p$ (all lower orders satisfy $v_n=0$ for $1<n<p$), contains both contributions from $V$ and $F$ and can be used as a good approximation to $u$ (assuming the series converges), what will be discussed in the next section. 

In this case we also have a convergence result
\begin{Theorem} \label{Th:V-Fu-pert2}
With $f,g, V$ and $F(u)$ as above for any $m>3$, $k>2$, $p> 1+\sqrt{2}$, $\la<C_{q,k}^{-1}$ and sufficiently small $\eps$ the series defined in \refa{Fu-V:pert-ser'}-\refa{V-Fu-2perteq} converges (in norm) in $\Linftx{q}$ for $q=\min(p-1,k,m-1)$ to the solution of the equation \refa{Fu-V:wave-eq} with initial data \refa{Fu-pert-initdata}.
\end{Theorem}
\begin{proof}
  The proof is again analogous to that of theorems \ref{Th:Fu-pert} and \ref{Th:V-Fu-pert1} with the following differences. We have for $q:=\min(m-1,k,p-1)$
\begin{equation*}
   \|v_1\|_\Linftx{q} \leq \|v_1\|_\Linftx{m-1} \leq C_m(f_0+f_1+g_0) =: D < \infty
\end{equation*}
and 
\begin{equation*}
\begin{split}
  \|v_{n+1}\|_\Linftx{q} 
  &\leq \|L_0(F_n(v_1,...,v_n))\|_\Linftx{q} + \law \|L_0(V v_{n+1-a})\|_\Linftx{q} \\
  &\leq  C \sum_k |a^n_k|\cdot \|v_1\|_\Linftx{q}^{\alpha^{n,1}_k} \cdots \|v_n\|_\Linftx{q}^{\alpha^{n,n}_k} + \w{\delta} \|v_{n+1-a}\|_\Linftx{q}.
\end{split}
\end{equation*}
with $\w{\delta}:=\law C_{q,k}$. The corresponding algebraic equation becomes now
\begin{equation} \label{algebraic-Fu-Va}
  {W = C\; \w{F}(W) + \w{\delta} \eps^a W + D \eps}.
\end{equation}
It cannot be rewritten, like before, as $G(W)=\eps$, but it can be written as
\begin{equation*}
  G(W,\eps):=W - C\; \w{F}(W) - \w{\delta} \eps^a W - D \eps = 0.
\end{equation*}
Since $a$ is a positive integer number, $G(W,\eps)$ is analytic in both variables around the point $G(0,0)=0$. Moreover, $\d G(W,0)/\d W|_{W=0} = 1-\w{\delta} \eps^a = 1 - \la C_{q,k}>0$. 
Then, by (real) analytic implicit function thorem (see e.g. \cite{Hille}), there exists a unique function $W(\eps)$ such that $G(W(\eps),\eps)=0$. Then $W(\eps)$ has a Taylor series representation with positive radius of convergence.
Repeating the reasoning of the previous proofs, we arrive at the conclusion that $\sum_{n=1}^\infty \eps^n \|v_n\|_\Linftx{q}$ has a positive radius of convergence $\w{R}>0$.
It follows that the series \refa{Fu-V:pert-ser'}-\refa{V-Fu-2perteq} converges in norm in $\Linftx{q}$ for all $0<\eps<\w{R}$ to the solution of \refa{Fu-V:wave-eq} with initial data \refa{Fu-pert-initdata}. 
Uniqueness follows again from theorem \ref{Th:V-Fu}.
\end{proof}

\subsection{Optimal decay estimate}

\subsubsection{Iteration with perturbative treatment of $V$}

From \refa{Fu-V:u-u_n} we have
\begin{equation*}
  |u(t,x)-u_n(t,x)| \leq \frac{\delta^n (2-\delta)^n}{(1-\delta)^2} 
  \cdot \frac{3C_m\eps}{\norm{t+|x|}\norm{t-|x|}^{q-1}}
  \leq \frac{c_n(x)\la^n\eps}{(1+t)^q}
\end{equation*}
for sufficiently small $\eps$ and $\la$ (such that $\la C_{q,k} = \delta<\delta_0<1$).
If we are able to show that some $u_n(t,x)\cong \eps d_n(x) t^{-q}\neq 0$ for large $t$ (the asymptotic approximation ``$\cong$'' is to be understood in the sense defined in \refa{def-approx}, with the relative error $\eta$) then already $u_n$ shows the correct decay rate, identical with this of $u$, because then, choosing $\eta:=\la^n$, we get
\begin{equation*}
  \left|u(t,x) - \frac{d_n(x)\eps}{t^q}\right| \leq 
  \frac{[d_n(x)+c_n(x)]\la^n\eps}{t^q}.
\end{equation*}
For small $\la$ it follows 
\begin{equation*}
  |u(t,x)| \cong \frac{d_n(x)\eps}{t^q}.
\end{equation*}
Again, in case when $m>p$, we have $u_1=I_0(f,g)\in\Linftx{m-1}$ and 
\begin{equation*}
  |u_1(t,x)|\leq \frac{C_m\cdot(f_0+f_1+g_0)}{\norm{t+|x|}\norm{t-|x|}^{m-2}} \leq
  \frac{2 C_m\cdot(f_0+f_1+g_0)}{(1+t)^{m-1}}
\end{equation*}
for $t\geq 2(m-2)|x|$,
hence it decays faster than $u$ and it cannot be shown that $u_1(t,x)\cong \eps d_1(x) t^{-q}$. It is expected that it will be true for $u_2\cong \eps d_2(x) t^{-q}$, what means, that already $u_2$ would have the same rate of decay as $u$ (see \cite{NS-PB_Tails} for such results in spherical symmetry).

In case when $m \leq  p$, we have $q:=m-1$. Then it should be possible to show $u_1(t,x)\cong d_1(x) t^{-(m-1)}$, what means, that already $u_1$ would have the same rate of decay as $u$.

\subsubsection{Perturbation series with perturbative treatment of $V$}

Consider the system \refa{V-Fu-2perteq-}-\refa{V-Fu-2perteq} and choose the constant $a:=p-1$ so that $v_2,...,v_{p-1}=0$ and at the order $v_p$ both effects, the nonlinear and linear (potential) scattering, appear simultaneously
\begin{align}
  v_{-n} &:= 0,\qquad n\geq 0\\
  v_1&=I_0(f,g) \label{Fu-V:pert-v1}\\
  v_2&=v_3=...=v_{p-1}=0\\
  v_{p} &= -\law L_0(V v_1) +  L_0(F_{p-1}(v_1,...,v_{p-1})) = -\law L_0(V v_1) +a_0 L_0((v_1)^p) \label{Fu-V:pert-vp}\\
  v_{n+1} &= -\law L_0(V v_{n-p+2}) +L_0(F_n(v_1,...,v_n)),\qquad n\geq p.
\end{align}
Consider only the more interesting case $m-1>\min(p-1,k)=:q$. If we can show that $v_p\cong d_p(x) t^{-q}$, then already $v_p$ has the correct decay rate, identical with this of $u$. To prove it, we can repeat the reasoning from the section when we treated nonlinear wave equation without the potential term, because the only fact, which we use is that the perturbation series $\sum_{n=1} \eps^n v_n$ has a positive radius of convergence and this is here guaranteed by theorem \ref{Th:V-Fu-pert2}. Analogously, we obtain
\begin{equation*}
  u(t,x) \cong \frac{d_p(x) \eps^p}{t^q}
\end{equation*}
for all $t>T$ and sufficiently big $T=T(\eps)$, so $v_p$ dominates the perturbation series for large times and small $\eps$ and has the same decay rate as the full solution of the nonlinear wave equation $u$.

This is the simplest setting for applications. Here, we only need to solve (approximately) two linear wave equations, \refa{Fu-V:pert-v1} and \refa{Fu-V:pert-vp}, in order to determine the decay rate for solutions of \refa{Fu-V:wave-eq} . This is the starting point of \cite{NS-PB_Tails} where we solve the two equations under spherical symmetry.


\appendix
\section{Some useful estimates}

The first two lemmas we cite from \cite{NS-WaveDecay}.

\begin{Lemma} \label{Lem:init-data}
Let the data $(f,g)\in\Linfx{m-1}\times\Linfx{m}$ with $m>3$ satisfy 
\begin{equation*}
  f_0:=\|f\|_\Linfx{m-1} < \infty,\qquad
  f_1:=\|\nabla f\|_\Linfx{m} < \infty,\qquad
  g_0:=\|g\|_\Linfx{m} < \infty.
\end{equation*}
Then there exists a unique weak solution $v(t,x)=I_0(f,g)$ of the free wave equation
\begin{equation*} 
  \Box v = 0,\qquad v(0,x)=f(x),\qquad \d_t v(0,x)=g(x)
\end{equation*}  
which satisfies
\begin{equation*}
  \|v\|_\Linftx{m-1} \leq C(f,g) := C_m\cdot (g_0+f_1+f_0).
\end{equation*}
\end{Lemma}

\begin{Lemma} \label{Lem:source}
Let the source $F$ satisfy for some $q>2$ and $1<p\leq q$
\begin{equation*}
  F_0:=\|\norm{x}^q F\|_\Linftx{p} < \infty.
\end{equation*}
Then there exists a weak solution $v(t,x)=L_0(F)$ of the free wave equation with source
\begin{equation*} 
  \Box v = F,
\end{equation*}
and null initial data $v(0,x)=0$, $\d_t v(0,x)=0$. Moreover, it satisfies
\begin{equation*}
  \|v\|_\Linftx{p} \leq C_{p,q} F_0.
\end{equation*}
\end{Lemma}

Next lemma we cite after Asakura \cite[Cor. 2.4 and Eq. 2.33]{Asakura} and state in our notation.
\begin{Lemma} \label{Lem:power-p}
Let $u\in\C^2(\RTX)\cap\Linftx{q}$ for some $q>1$. Then for any $p>1+\sqrt{2}$
\begin{equation*}
  \|L_0(|u|^p)\|_\Linftx{q} \leq C \|u\|_\Linftx{q}^p
\end{equation*}
with some $C>0$ provided $q\leq p-1$.
\end{Lemma}
Note, that it is a consequence of lemma \ref{Lem:source}, but only when $p>3$, while for $1+\sqrt{2}<p\leq 3$ it requires a more general proof. It can be easily deduced, also for weak solutions $u\in\C^0(\RTX)$, from a more general estimate \cite{NS-DecayLemma}
\begin{Lemma} \label{Lem:decay}
If 
\begin{equation*} 
   |F(t,x)|\leq \frac{A}{\norm{t+|x|}^p\norm{t-|x|}^q}
\end{equation*}
with $p>2$, $q>1$ then 
\begin{equation*} 
   |L_0(F)(t,x)|\leq \frac{C}{\norm{t+|x|}\norm{t-|x|}^{p-2}}
\end{equation*}
with some positive constant $C$.
\end{Lemma}

\bibliography{QNMs}
\bibliographystyle{unsrt}

\end{document}